\documentclass[pra,aps,twocolumn,showpacs]{revtex4}
\usepackage{amsmath,graphicx,epsfig}
\usepackage{color}
\usepackage{amssymb,amsmath,array,graphicx,epsfig,paralist}
\usepackage{multirow}
\usepackage{booktabs}
\usepackage{setspace}

\def\fig_width{3.375 in} % width of single column figure in PR
\def\fig_width{3. in} % width of single column figure in PR
\newcommand{\tab}{\hspace*{2em}}
\begin{document}
\title{Measurement of collisions between rubidium atoms and optically dark rubidium ions in trapped mixtures}
%==================================
%
\author{Seunghyun Lee}
\email{lee@rri.res.in}
\author{K. Ravi}
\author{S. A. Rangwala}
\email{sarangwala@rri.res.in}
\affiliation{Raman Research Institute, Sadashivanagar, Bangalore 560080, India}
\date{\today}
%%%----------------------------------------------------------------------

\begin{abstract}
We measure the collision rate coefficient between laser cooled Rubidium (Rb) atoms in a magneto-optical trap (MOT) and optically dark Rb$^+$ ions in an overlapping Paul trap. In such a mixture, the ions are created from the MOT atoms and allowed to accumulate in the ion trap, which results in a significant reduction in the number of steady state MOT atoms. A theoretical rate equation model is developed to describe the evolution of the MOT atom number, due to ionization and ion-atom collision, and derive an expression for the ion-atom collision rate coefficient. The loss of MOT atoms is studied systematically, by sequentially switching on the various mechanisms in the  experiment. Combining the measurements with the model allows the direct determination of the ion-atom collision rate coefficient. Finally the scope of the experimental technique developed here is discussed.
\end{abstract}
\maketitle
\section{Introduction}
\tab Trapped ion and atom mixtures enable us to investigate exciting physics ranging from,
a single trapped ion inside a Bose-Einstein Condensate (BEC)~\cite{Zip10, Sch10} and Magneto-Optical Trap (MOT) atoms superimposed with laser cooled ions~\cite{Gri09, Fel11, Wad11}, to mixtures where the ions are only cooled in collision with the atoms~\cite{Rav12, Sco11, Arn12}. Elastic and charge exchange processes in these mixed, trapped system have been demonstrated and measured~\cite{Gri09,Fel12, Arn12}. Even before the existence of such mixed traps, a variety  of rate coefficients and cross-sections for several ion-atom processes have been measured~\cite{Chu93}. The most prevalent method for the determination of the ion-neutral collision rate using trapped ions, measures the loss of ions from the trap~\cite{Chu93, Dra05}. For optically dark ions, this loss is measured by sweeping the ion resonance frequency or by extracting the trapped ions to charge particle detectors~\cite{Chu93}. However, in mixed ion-atom trapping experiments in which the ions are optically dark, the cold trapped atoms interacting with ions can be directly measured in situ, by atomic fluorescence, without affecting the trapped ions. In this article, we develop a method to measure ion-atom collision rate coefficient in such a hybrid system using atom fluorescence as a probe.

\tab In the experiment reported here, we trap the optically dark Rubidium ($^{85}$Rb$^+$) ions, derived from the laser cooled $^{85}$Rb atoms in a MOT. The ions are created and accumulated at the center of the ion trap, which overlaps with the MOT. Such a system evolves into a mixture with constant numbers of trapped atoms and ions  in steady state. The ionization of the MOT atoms and the subsequent trapped ion interaction with the atoms results in a depletion of the atoms from the MOT. The consequent drop in MOT fluorescence is utilized to develop the framework for the measurement of the collision rate coefficient for ion-atom interactions, as described below.

\tab In what follows, we construct the method for the determination of the collision rate coefficient. We first present a brief description of the experimental setup, followed by a simple, analytical rate equation model, where we systematically consider loading and loss processes of the MOT atoms. Here we model in a calibrated manner the change in the trapped atom number when each ion related process is turned on. This culminates with the simultaneous trapping of ions and atoms in steady state. The experimental data is then presented in a sequence consistent with the processes modeled. Combining the measurements with the derived analytical expressions allows the determination of the rate of atom loss due to ion-atom interactions, and hence the collision rate coefficient for ions and atoms. We end with a discussion of some features of such measurements.

\section{Experimental Setup}
%----------------------------------------------------------------------------
\begin{figure}
\includegraphics[width=8.3 cm]{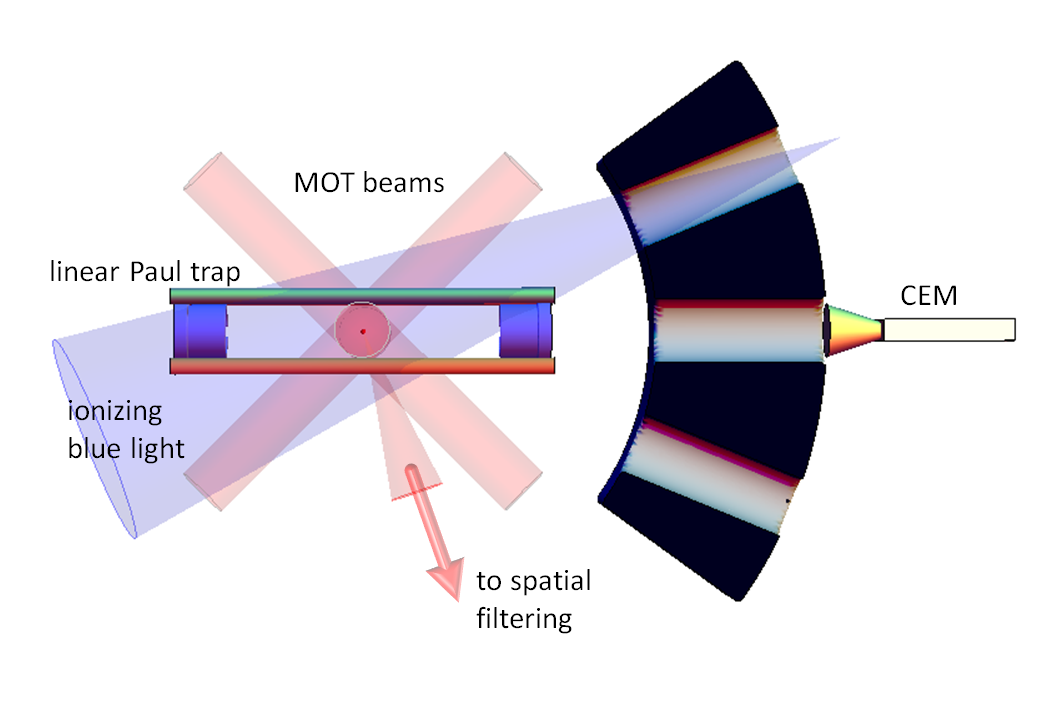}
\caption{(color online) The schematic diagram of the experimental setup is illustrated. The MOT and the linear Paul trap are overlapped as shown, with the cooling lasers illustrated in red, and the MOT located at the intersection point represented by a red sphere. The ionizing blue light is illustrated, without its source. The channel electron multiplier (CEM) is used to detect the trapped ions, by appropriately switching the voltage on the hollow end cap electrode closer to the CEM. The direction for the MOT fluorescence detection is indicated, without illustrating the spatial filtering arrangement.}
\label{Fig:chamber}
\end{figure}
%----------------------------------------------------------------------------
\tab The experimental arrangement (Fig.~\ref{Fig:chamber}), techniques and characterization have been described in detail in previous work~\cite{Rav11}. Here we include a brief description for making the experimental discussion comprehensible. Rb atom vapor is created from a dispenser and $^{85}$Rb atoms are laser cooled and trapped in a six-beam MOT of size = 0.35 mm (standard deviation) and temperature = 166($\pm$ 28) $\mu$K. The cooling laser is red detuned by  $\delta_c= 2 \pi \times 10$ MHz with respect to the $5s_{1/2}(F=3)\leftrightarrow5p_{3/2}(F'=4)$ transition, and the laser power is 3.5 mW, per beam. A magnetic gradient field of 12 Gauss/cm is utilized for the MOT. The intersection point of the MOT laser intensity maxima and magnetic field zero coincides with the center of a linear Paul trap, built around the capture volume of the MOT. This ensures that the trapped ions and laser cooled atoms are optimally overlapped~\cite{Rav12, Rav11}. Ions are created from cold atoms using a blue light source (BLS) which along with the cooling light results in two photon ionization of the Rb MOT atoms. The ion trap is made up of four parallel rods in the quadrupole configuration, where a time varying radio-frequency (RF) voltage is applied to one diagonal set, and an opposite phase RF voltage to the complementary diagonal set. This generates a two dimensional quadrupole field for the radial trapping of ions. Two end cap ring electrodes at a positive dc voltage confine the positive ions along the trap axis. Atoms are detected by observing the fluorescence from the MOT on a photo-multiplier tube (PMT) and imaged using two CCD cameras. The MOT fluorescence is measured by the PMT, where a spatial filtering arrangement is used to reduce background light signal. The ions are optically dark and are therefore detected either indirectly by a change in atomic fluorescence of the MOT or destructively by extraction onto a channel electron multiplier (CEM) and measuring the ion counts, either by pulse counting or by measurement of the proportional extracted ion signal. The ion extraction to the CEM is affected by switching the ion trap end cap electrode from $+80$ V to $-5$ V, while the RF field is still on.
%%%%%%%%%%%%%%%%%%%%%%%%%%%%%%%%%%%%%%%%%%%%%%%%%%%%%%%%%%%%%%%%%%%%%%%%%%%%%%%%%%%%%%%%%%%%%%%%%%%%%%%%%%%%%%%%%%%%%%%%%%%%%%%%%%%%%%%%%%%%
\section{Model for the Loss of Trapped Atoms}
\subsection{The MOT rate equation}
\tab We now present the rate equation framework which describes the evolution of the ion atom system as we add the various ingredients of the above experiment one at a time. The total number of atoms in the MOT ($N_{MOT}$) at any given time, is $N_{MOT} = N_{g} + N_{e}$, where $N_{g}$ and $N_{e}$ are the atom numbers in the ground and excited states respectively. The time dependence of $N_{MOT}$ can be written as 
\begin{equation}
\label{eq:MOT_rate}
\frac{dN_{MOT}}{dt}= L - \gamma_{ml} N_{MOT},
\end{equation}
with $L$, the loading rate of the atoms from the background vapor into the MOT and $\gamma_{ml}$, the loss rate of the MOT atoms, which is primarily collisional ~\cite{Tow95}. The time dependent solution of eqn~(\ref{eq:MOT_rate}) with initial condition $N_{MOT}(0)=0$ is
\begin{equation}
\label{eq:MOT_sol}
N_{MOT}(t)= N_0 (1 - e^{-\gamma_{ml} t}),~~ N_0 = \frac{L}{\gamma_{ml}}.
\end{equation}
where $N_0$ is the steady state atom number in the MOT, in the limit of large time.

\subsection{Fraction of excited atoms}
\tab The fraction of atoms in the excited state $f_{e}$ is determined from the photon scattering rate from an atom and can be written as ~\cite{Ste12}
\begin{equation}
\label{eq:exc_population}
f_{e} = \frac{N_{e}}{N_{g} + N_{e}} = \frac{{(\Omega / \Gamma})^2}{1 + 4 (\delta_{c} / \Gamma)^2 + 2 (\Omega / \Gamma)^2},
\end{equation}
where, $\Omega = 2 \pi \times 16.7$ MHz is the Rabi frequency and $\Gamma =2 \pi \times 6$ MHz is the decay rate of the excited atomic state. 

\subsection{Photo ionization of atoms}
\tab The ionization of atoms in our experiment is accomplished by excitation of the atoms in the $5p_{3/2}$ state to the ionization continuum using the BLS. The central wavelength the BLS is at $\lambda_{pi} = 456$ nm, which is sufficient to ionize the excited Rb atoms~\cite{San05}. The intensity of the BLS light incident on the MOT is denoted by $I_{pi}$. For photo ionization energy $E_{pi} = h c/ \lambda_{pi}$, the ionization cross-section is denoted by $\sigma_{pi}$. The photon energy which is in excess of the ionization energy is carried away by the ejected electron, with the resulting ion suffering the recoil. 

\tab The loading rate of the MOT remains unchanged when the BLS is switched on for photo ionization, as the population of the background atoms remains unchanged with operation of the BLS. The maximum intensity of the BLS used is 1.33 mW/cm$^2$. The rate equation for the MOT, when the BLS is ON is modified to
\begin{equation}
\label{eq:PI_rate}
\frac{dN_{MOT}}{dt}= L - \gamma_{ml} N_{MOT} - \gamma_{pi} N_{MOT},
\end{equation}
where a new loss rate $\gamma_{pi}$ adds to the losses inherent to the operation of the isolated MOT to give a total loss rate of $\gamma_{t}= \gamma_{ml}+\gamma_{pi}$~\cite{Gab97, Fus00}. Here we assume that since the ion trap is switched off during photo ionization, the ions and electrons created from the MOT atoms immediately leave the system and play no further role in determining the number of trapped atoms. The ionization rate of the excited Rb atoms by radiation from the BLS can be written as~\cite{Gab97, Fus00}
\begin{equation}
\label{eq:loss_PI}
\gamma_{pi} = \left(\frac{\sigma_{pi} f_{e}}{E_{pi}}\right) I_{pi} = \zeta I_{pi},
\end{equation}
where $\zeta = \left(\frac{\sigma_{pi} \lambda_{pi} f_{e}}{h c}\right)$.
The corresponding time dependence of $N_{MOT}$ with initial condition $N_{MOT}(0)=0$ and the inclusion of the ionizing process can then be written as
\begin{equation}
\label{eq:PI_sol}
N_{MOT} (t) = \frac{L}{\gamma_{t}} (1 - e^{-\gamma_{t} t}).
\end{equation}

\subsection{Ion-Atom interaction loss}
\tab Once the ion trap is switched ON, then a fraction of the created ions load into the ion trap which overlaps the MOT. Since the ion and atom traps are optimally overlapped, the ions created from the MOT have negligible velocity upon creation. These ions gain in kinetic energy as they evolve in the field of the ion trap. As the number of ions loaded in the ion trap grows, the spatial distribution of the trapped ions increases outward from the trap centers. Since the MOT volume is much smaller than the volume of the trapped ion spatial distribution, therefore long before the total number of trapped ions equilibrates, the number of ions at any instant overlapped with the MOT stabilizes. Therefore the collisions rate between the trapped ions and the MOT atoms also stabilizes rapidly, allowing the definition of a time independent binary ion-atom interaction rate, $\gamma_{ia}$ for the constituents of the ion-atom mixture. 
Since atoms are weakly trapped with respect to ions, they can gain sufficient energy in collision with trapped ions to exit the MOT. This is a new loss channel for the atoms trapped in a MOT, which can be written as,
\begin{equation}
\label{eq:IA_intl}
\frac{dN_{MOT}}{dt}= L - (\gamma_{ml} + \gamma_{pi} + \gamma_{ia}) N_{MOT}.
\end{equation}
The resulting time evolution for $N_{MOT}$ with initial condition $N_{MOT}(0)=0$ therefore becomes,
\begin{equation}
\label{eq:IA_sol}
N_{MOT} (t) = \frac{L}{\gamma_{tot}} (1 - e^{-\gamma_{tot} t}),
\end{equation}
where $\gamma_{tot} = \gamma_{t} + \gamma_{ia}$ is the total loss rate of MOT atoms, when the ion and atom traps are operated simultaneously as described in the experimental section above. We note that while the ion density overlapping with MOT volume stabilizes rapidly, the number and the velocity distribution of the trapped ions evolve continuously till the ion trap is filled to its limit. Thus the rate of ion loss from the ion trap and the evolution of $N_{MOT}$ are linked.

\subsection{Number of trapped ions}
\tab The ion trap is loaded to the minimum of the secular trap potential, by photoionization of the overlapping MOT atoms. The number of trapped ions as a function of time, for a given photo-ionizing intensity $I_{pi}$, can be described as
\begin{equation}
\label{eq:IT_time}
N_{I}(t) = N_{MOT} \left.\zeta\right. \frac{I_{pi}}{\gamma} (1 - e^{-\gamma t}),
\end{equation}
where $\gamma$ is the ion trap loss rate. Since ions have much higher velocities than the MOT capture velocity for the atoms, ion-atom collisions will lead to loss of atoms from the MOT. This allows the number of MOT atoms lost due to trapped ions $N_{MOT}^{loss}$, to be written as
\begin{equation}
\label{eq:MOT_loss}
N_{MOT}^{loss}(t) = N^{'} (1 - e^{-\gamma_{ia} t}), 
\end{equation}
where $N^{'}$ is saturation number of atom loss due to ion-atom interaction, and $\gamma_{ia}$ is the loss rate of the MOT atoms due to ion-atom interactions. The above form of the atom loss will be validated later (eqns.~(\ref{eq:PI_newsol}),(\ref{eq:IA_newsol}) and ~Fig.~\ref{Fig:sequence}), when the experiment and results are presented. The atom loss rate from the MOT, due to ion atom interactions, is proportional to the number of trapped ions ($N_{MOT}^{loss}(t) \propto N_{I}(t)$), and so from eqns.~(\ref{eq:IT_time}) and~(\ref{eq:MOT_loss}) we conclude that $\gamma \equiv \gamma_{ia}$. Therefore, number of trapped ions as a function of t is expressed as,
\begin{equation}
\label{eq:IT_time_sol}
N_{I}(t) = N_{MOT}\left.\zeta\right. \frac{I_{pi}}{\gamma_{ia}} (1 - e^{-\gamma_{ia} t}).
\end{equation}
The number of trapped ions also depends on $I_{pi}$ but cannot increase indefinitely with $I_{pi}$ due to finite trap depth and volume. Therefore an intensity loss coefficient of trapped ions must be introduced. In the same spirit as the above discussion, to construct a general form of $N_{I}$, we write
\begin{equation}
\label{eq:IT_int}
\frac{dN_{I}(t, I_{pi})}{dI_{pi}} = N_{MOT} \left.\zeta\right. \frac{1}{\gamma_{ia}} (1 - e^{-\gamma_{ia} t}) - \kappa N_{I}(t, I_{pi}),
\end{equation}
where $\kappa$ is the intensity loss coefficient and has the unit of inverse intensity.
The solution of the above equation is, then
\begin{equation}
\label{eq:IT_gen_sol}
N_I(t, I_{pi}) = N_{MOT} \left.\zeta\right. \frac{1}{\gamma_{ia} \kappa} (1- e^{-\gamma_{ia} t})(1 - e^{-\kappa I_{pi}})
\end{equation}
The equation can be fully converted into a time dependent function of trapped ions when $I_{pi}\rightarrow\infty$ and also fully converted into an $I_{pi}$ dependent function of trapped ions when $t\rightarrow\infty$. The number of ions trapped as a function of $I_{pi}$ for $t\rightarrow\infty$ is written as
\begin{equation}
\label{eq:IT_sol}
N_{I}(I_{pi}) = N_{I}^{0} (1 - e^{- \kappa I_{pi}}).
\end{equation}
The maximum trapped ion number, $N_{I}^{0}$ when $t\rightarrow\infty$ and $I_{pi}\rightarrow\infty$ is expressed as,
\begin{equation}
\label{eq:IT_steady_sol}
N_{I}^0 = N_{MOT}\left.\zeta\right.\frac{1}{\gamma_{ia} \kappa}.
\end{equation}
Since eqns.~(\ref{eq:IT_sol}) and~(\ref{eq:IT_steady_sol}) apply to the ions system in the limit $t\rightarrow\infty$, it is therefore reasonable to expect an average motional energy $\left\langle E_{I}\right\rangle$, for each ion.

\subsection{Determination of ion-atom collision rate}

\tab The ions trapped as described above are assumed to have a speed distribution $f(v)$, consistent with $\left\langle E_{I}\right\rangle$. The laser cooled atoms on the other hand are expected to exhibit a temperature $T_{A}\approx100~\mu$K. In this situation $\left\langle E_{I}\right\rangle \gg k_B T_{A}$ and therefore all of the velocity of the ion-atom collision can be assumed to be possessed by the ion, in the lab frame of reference (LFoR). Similarly, as is shown later, $N_{MOT}\gg N_{I}^{0}$ as the ion density $\rho_{I}$, is far less that the atom density $\rho_{A}$ in such mixtures despite the ion trap volume, $V_{IT} \gg V_{MOT}$, the MOT volume in our experiment.

\tab The total ion-atom collision cross-section, $\sigma_{tot}$ is energy ($E\propto v^2$) dependent and is the sum of the elastic, $\sigma_{el}$, and resonant charge exchange, $\sigma_{cx}$. $\sigma_{el}\propto 1/{E^{1/3}}$ over all the energies and  $\sigma_{cx}\propto 1/E^{1/2}$ in the low energy regime and $\sigma_{cx}\propto (a$ ln $E - b )^2$ at high collision energies ($a$ and $b$ are constants)~\cite{Cot00}. Given $f(v)$ for the ions, the determination of
$\sigma_{tot}$ rests on assumptions made for $f(v)$. In this case the experimentally meaningful quantity to consider is the ion-atom rate coefficient, 
%----------------------------------------------------
\begin{equation}
\label{eq:ratecoeff}
k_{ia} = \int_{v} \sigma_{tot}\left.v\right.f(v)dv \equiv \left\langle  \sigma_{tot}\left.v\right.\right\rangle,
\end{equation}
%-----------------------------------------------------
which represents the velocity averaged product of $\sigma_{tot}v$. A single ion with velocity $v_0$ and a corresponding ion-atom cross-section $\sigma^0_{tot}$ will collide with MOT atoms of density $\rho_{A}$ at a rate given by  
%-----------------------------------------------
\begin{equation}
z_0 = \sigma^0_{tot} \left.v_{0}\right. \rho_{A}.
\end{equation}
%-----------------------------------------------
For the $N_{I;M}$ ions that overlap the MOT at any given time, with a speed distribution $f(v)$, the total ion-atom collision rate then becomes,  
%-----------------------------------------------
\begin{equation}
\label{eq:collrateIA}
z N_{I;M} = \left\langle  \sigma_{tot}\left.v\right. \right\rangle \rho_{A} N_{I;M} \equiv k_{ia}  \left.\rho_{A}\right. N_{I;M}.
\end{equation}
%-----------------------------------------------
For a given value of $I_{pi}$, the number of ions in the ion trap volume $V_{IT}$, overlapping the MOT volume $V_{MOT}$, is given by,
%-----------------------------------------------
\begin{equation}
\label{eq:overlap_ion}
N_{I;M} = N_{I}^{0} (1 - e^{-\kappa I_{pi}})\frac{V_{MOT}}{V_{IT}}.
\end{equation}
%-----------------------------------------------
\tab Since ion velocities are large, and the MOT capture velocity for the atoms much smaller, ion-atom collisions cause the MOT atoms to eject, allowing us to equate the total ion-atom collision rate in eqn.~(\ref{eq:collrateIA}) to $\gamma_{ia} N_{MOT}$, the atom loss rate due to ion-atom collisions. After substituting eqn.~(\ref{eq:overlap_ion}) into eqn.~(\ref{eq:collrateIA}), the resulting expression for $\gamma_{ia}$ then becomes,
%-----------------------------------------------
\begin{equation}
\label{eq:loss_IA_interm}
\gamma_{ia} = \frac{N_{I}^{0}k_{ia}}{V_{IT}} (1 - e^{-\kappa I_{pi}}),
\end{equation}
%----------------------------------------------
which  defines the relation between the experimentally measurable $\gamma_{ia}$ and the rate coefficient for ion-atom collisions. In the experimental results that follow, we demonstrate that the above rate equation formulation describes adequately the dependence of the various loss rates with $I_{pi}$, the BLS intensity.

%%%%%%%%%%%%%%%%%%%%%%%%%%%%%%%%%%%%%%%%%%%%%%%%%%%%%%%%%%%%%%%%%%%%%%%%%%%%%%%%%%%%%%%%%%%%%%%%%%%%%%%%%%%%%%%%%%%%%%%%%%%%%%%%%%%%%%%%%%%%%%
%
\section{Results}

\subsection{Experimental Sequence}

%----------------------------------------------------------------
\begin{figure}
\includegraphics[width=8.3 cm]{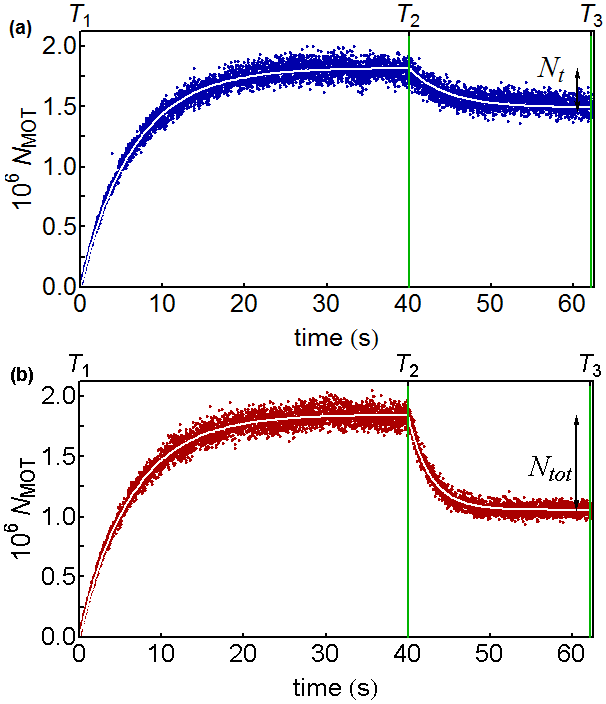}
\caption{(color online) The temporal sequence for the experiment is illustrated. In both panels, first the MOT is loaded for 40 s (from T$_1$ and T$_2$) to saturation, where it contains $\approx 1.8\times 10^6$ atoms. Beyond T$_2$, the photo ionization light is switched on. Panels (a) and (b) illustrate the evolution of the trapped atoms when the ion trap is OFF and with ion trap ON, respectively. The number of trapped atoms reduces in both cases. However, when the ions and the atoms are simultaneously trapped, the atom loss is much more significant. The plots in (a) and (b) correspond to BLS intensity $I_{pi} = 1.33$ mW/cm$^2$. The various loss rates discussed in the text are determined from such data as a function of photo-ionizing intensity. $N_t$ and $N_{tot}$ are the atom losses due to $\gamma_t$ and $\gamma_{tot}$. The lines show the fits to the data according to eqns. (\ref{eq:PI_newsol}) and (~\ref{eq:IA_newsol}) repectively.}
\label{Fig:sequence}
\end{figure}
%----------------------------------------------------------------
\tab We now describe the experiment which allows us to validate the above rate equation analysis for ion-atom interaction. The basic time sequence instrumental for results below is shown in Fig.~\ref{Fig:sequence}. Here a MOT containing $\approx 1.80(\pm 0.06) \times 10^6$ atoms is loaded to saturation in 40 s (T$_1$ to T$_2$), and the change in the number of atoms in the MOT is measured under two conditions from T$_2$ to T$_3$ at different intensities of the ionizing light (BLS). When the BLS ionizes the cooled $^{85}$Rb atoms from the MOT and these ions and electrons leave the MOT volume immediately as there is no confinement for them. This results in a depletion of the steady state atom number in the MOT and therefore allows the determination of the ionization rate for the excited atoms. Using eqn.~(\ref{eq:MOT_sol}), data from T$_1$ to T$_2$ are fitted to determined $\gamma_{ml}$. Solution for eqn.~(\ref{eq:PI_rate}) with appropriate initial conditions, $N(t=0)=N_{0}$ and $N(t\rightarrow\infty)=N_{0}-N_{t}$ is
\begin{equation}
\label{eq:PI_newsol}
N_{MOT}(t)=N_{0}-N_{t}(1-e^{-\gamma_{t} t}),
\end{equation}

where $N_{t}$ is the atom loss due to $\gamma_t$ during photo ionization. $\gamma_t$ is determined by fitting data from $T_2$ to $T_3$ in Fig.~\ref{Fig:sequence}(a) with eqn.~(\ref{eq:PI_newsol}). In Fig.~\ref{Fig:sequence}(b) we switch ON both the ionizing light (BLS) and the ion trap simultaneously. In this case the ionization process is unaffected, but some fraction of the ions created reside within the ion trap, which is well overlapped with the atom trap. Here we observe that the number of trapped atoms reduces significantly over the ionization loss. This loss is attributed to ion-atom interactions. Solution for eqn.~(\ref{eq:IA_intl}) is obtained in a similar way as photo ionization case and given below.
%--------------------------------------
\begin{equation}
\label{eq:IA_newsol}
N_{MOT}=N_{0}-N_{tot}(1-e^{-\gamma_{tot} t})
\end{equation}
%---------------------------------------
where $N_{tot}$ is the atom loss due to $\gamma_{tot}$ when both atoms and ions are simultaneously trapped.

\subsection{Ionization without trapping}

%----------------------------------------------------------------
\begin{figure}
\includegraphics[width=8.3 cm]{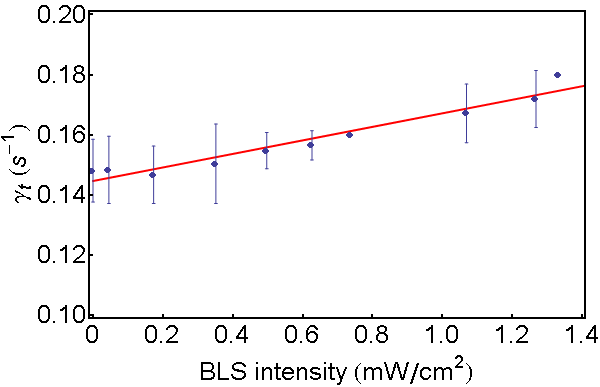}
\caption{(color online) The loss rate of atoms from the MOT as a function of photo ionization light intensity, $I_{pi}$, without the ion trap in Fig.~\ref{Fig:sequence} (a). As $I_{pi}$ increases, the loss of atoms from the MOT increases as does the rate of loss $\gamma_{pi}$. Since this atoms loss is happening on top of the normal loss from the MOT, the experimentally measured quantity here is $\gamma_{t}$. As can be seen from the text, a linear relation is expected between $\gamma_{t}$ and $I_{pi}$, which is measured with a slope of $0.00225 (\pm 0.0002)$ m$^2$/J and the $\gamma_t$ axis intercept of $\approx 0.14$ s$^{-1}$ which is the MOT loss rate without the ionizing light (BLS). Error on the slope from the fit is calculated from confidence level analysis.}
\label{Fig:loss_BLS}
\end{figure}
%----------------------------------------------------------------
\tab When the BLS is switched ON at T$_2$, and the ion trap is not operational, the loss rate of atoms from the MOT increases due to ionization of the atoms, which is shown in Fig.~\ref{Fig:loss_BLS}. The measured loss rate coefficient $\gamma_{t} = \gamma_{ml} + \gamma_{pi}$ can be written using eqn.~(\ref{eq:loss_PI}) as, 
%----------------------------------
\begin{equation}
\gamma_{t} = \gamma_{ml} + \zeta I_{pi}
\label{eq:loss_t_fit}
\end{equation} 
%---------------------------------
From Fig.~\ref{Fig:loss_BLS} it is clear that $\gamma_{t}$ has a linear dependence on the intensity of the BLS, $I_{pi}$. This is because although the ionization process is a two photon process, the operation of the MOT ensures that a constant fraction of atoms are present in the excited state, making the ionization an effective single photon process. The slope of $\gamma_t$, $\zeta$ is determined to be 0.00225$(\pm 0.0002)$ m$^2$/J from the data. The value of $\gamma_t$ for $I_{pi} = 0$ is $\gamma_{ml}=0.14 (\pm 0.011)$ s$^{-1}$.

\subsection{Ionization and ion trapping}
%----------------------------------------------------------------
\begin{figure}
\includegraphics[width=8.3 cm]{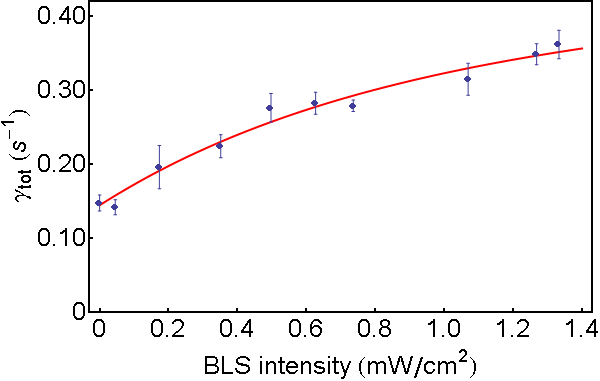}
\caption{(color online) The loss rate of atoms from the MOT as a function of ionization light intensity, $I_{pi}$, with the ion trap in Fig.~\ref{Fig:sequence} (b). As $I_{pi}$ increases, the loss of atoms occurs on top of the normal loss from the MOT and the ionization loss, and so the experimentally measured quantity here is $\gamma_{tot}$. The additional loss of atoms over that in  Fig.~\ref{Fig:loss_BLS} is attributed to the ion-atom interactions for which the loss rate is $\gamma_{ia}$. $\gamma_{tot}$ has exponential and linear components in $I_{pi}$ as is observed.}
\label{Fig:loss_IT}
\end{figure}
\begin{figure}
\includegraphics[width=8.3 cm]{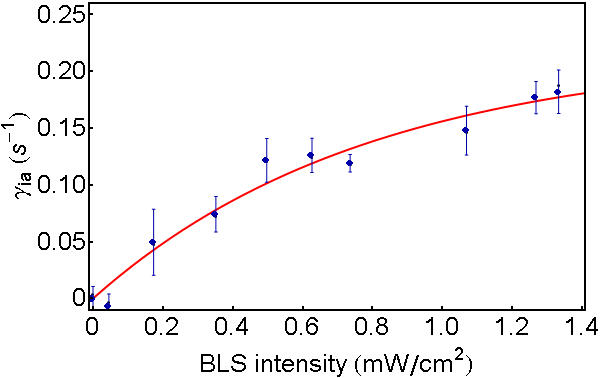}
\caption{(color online) A subtraction of the data in Fig.~\ref{Fig:loss_BLS} from that in Fig.~\ref{Fig:loss_IT} is plotted to isolate the ion-atom interaction term. This loss rate of the MOT atoms due to ion atom interaction, $\gamma_{ia}$ is plotted against BLS intensity, $I_{pi}$. $\gamma_{ia}$ is fitted with a single exponential function of the form, eqn.~(\ref{eq:loss_IA_interm}) and the coefficients of the fit are utilized to characterize the interaction term, as discussed in the text.}
\label{Fig:loss_IA}
\end{figure}

%----------------------------------------------------------------
\tab For the case when ions are held in the ion trap after T$_2$ in Fig.~\ref{Fig:sequence}(b), loss rate coefficient $\gamma_{tot}$ varies with a nonlinear behavior as shown in Fig.~\ref{Fig:loss_IT}. The change in $\gamma_{tot}$ is fit by a combination of the linear expression in eqn.~(\ref{eq:loss_t_fit}) and the exponential form of  eqn.~(\ref{eq:loss_IA_interm}). To study the loss rate due to ion-atom interaction, $\gamma_{ia}$ is plotted separately, as $\gamma_{ia} = \gamma_{tot} - \gamma_{t}$, in Fig.~\ref{Fig:loss_IA}. The exponential nature of the $\gamma_{ia}$ as a function of $I_{pi}$ is evident thus validating the rate equation for the ion-atom interaction. From the fit to the data eqn.~(\ref{eq:loss_IA_interm}) becomes, $\gamma_{ia}= 0.218 (1 - exp[-0.126~I_{pi}])$. Identifying  $\kappa = 0.126 (\pm 0.031)$ in inverse intensity units, i.e. m$^2$/W and 
%------------------------------------------
\begin{equation}
\label{eq:ion_atom_int_param}
\frac{N_{I}^{0} k_{ia}} {V_{IT}}=  0.218 (\pm 0.029)
\end{equation}
%-----------------------------------------
with units s$^{-1}$, we can constrain the product of the trapped ion density and the ion-atom collision rate coefficient.

\subsection{Direct ion detection}
%----------------------------------------------------------------
\begin{figure}[b]
\includegraphics[width=8.3 cm]{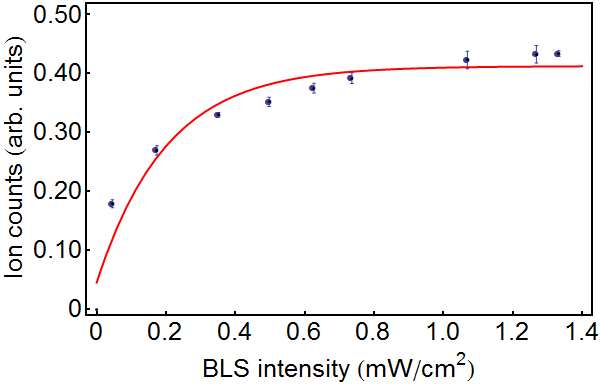}
\caption{(color online) The change in the detected ion counts vs. BLS intensity is shown. Since the number of trapped ions is very large, severe pileup of the ion signal results and the ion signal is seen to saturate faster than measured from the atom signal. The fit is generated using eqn.~(\ref{eq:IT_sol}).}
\label{Fig:iontrap}
\end{figure}

%----------------------------------------------------------------
\tab Alongside the measurements on the MOT discussed above, the trapped ions though optically dark in the present experiment, can be measured using a CEM as described in the experimental arrangement. At the end of each ion-atom experiment, the trapped ions are extracted into the CEM, by switching the end cap voltage appropriately. However, since this experiment fills the ion trap to its capacity, severe pile-up results due to overlapping arrival times of the ions onto the detector. For a CEM cone voltage of -2100 V, the extracted ions from the trap as a function of $I_{pi}$ is illustrated in Fig.~\ref{Fig:iontrap}. A single exponential dependence of the ion numbers detected with BLS intensity can be expected and a dependence of that nature is seen in the data. However there is detector saturation to contend with and the present measurement of ion numbers is limited by the detection. The direct measurement of the CEM signal is therefore presented for the sake of experimental completeness rather than with the motive of utilizing it to determine the ion-atom interaction rates. For large ion numbers this is likely to remain a significant problem with present day ion detectors. Alternative schemes for ion detections by electrode pickup, Faraday cup detection, etc.~\cite{Maj05}, may be useful for this problem and should be explored.  
%----------------------------------------------------------------
%----------------------------------------------------------------

\subsection{Collision rate coefficient determination}
%%---------------------------------------------------------------------
\begin{table*}[t]
\caption{The key values and errors for the quantities which are used in the determination of $k_{ia}$} % title of Table
\centering % used for centering table
\begin{tabular}{l l l l l l l} % centered columns (4 columns)
\hline\hline %inserts double horizontal lines
Parameter													&			& Value ($\pm$ error)           &   	& Units    & &   Method\\ [0.5ex] % inserts table
%heading
\hline % inserts single horizontal line
% inserting body of the table
$\zeta  $  														 & &$0.00225 (\pm 0.0002)$ 						  & &	 m$^2$/J  & &fit from the experimental data, in Fig.~\ref{Fig:loss_BLS} \\
$N_{MOT}$															 & & $1.80 (\pm 0.06)\times	10^{6}$		 	& &	 number   & &fluorescence measurement\\
$V_{IT}$ 															 & &$8.33 (\pm 0.83)\times 10^{-8}$			& &  m$^3$    & &derived from secular frequencies and trap depth  \\
$N_{I}^{0}$ 													 & &$1.48 (\pm 0.43)\times 10^{5}$	  	& &	 number   & &refer eq.~\ref{eq:IT_steady_sol}\\ 
$\gamma_{ia}(I_{pi}\rightarrow\infty)$ & & $0.218	(\pm 0.029)$  							& &	 s$^{-1}$ & &fit from the experimental data, in Fig.~\ref{Fig:loss_IA}\\
$\kappa$												       & & $0.126 (\pm 0.031)$							  & &	 m$^2$/W  & &fit from the experimental data, in Fig.~\ref{Fig:loss_IA} \\
$k_{ia}$  													   & &$1.23 (\pm 0.42)\times 10^{-13}$    & &  m$^{3}$/s & &calculated from the various quantities given above \\[1ex] % [1ex] adds vertical space
\hline %inserts single line
\end{tabular}
\label{table:error} % is used to refer this table in the text
\end{table*}
%%------------------------------------------------------------------
\tab We now demonstrate the use of eqn.~(\ref{eq:ion_atom_int_param}), to arrive at the ion-atom collision rate in the present experiment. The values of all the quantities used in the determination of the collision rate, with the standard deviation error in the accompanying parenthesis are provided in the text below. The product of the ion density and $k_{ia}$ is determined from the fit to eqn.~(\ref{eq:loss_IA_interm}) as $0.218 (\pm 0.029)$ /s, so only the determination of the number of trapped ions and the volume of trapping for the ions is required for computation of the collision rate coefficient. Keeping the compromised CEM data in mind, $\kappa = 0.126 (\pm 0.031)$ m$^{2}$/W is determined from the experimental data for atom loss Fig.~\ref{Fig:loss_IA}, rather than from the data in Fig.~\ref{Fig:iontrap}. Using eqn.~(\ref{eq:IT_steady_sol}), the maximum number of ions which can be accumulated in the ion trap volume is calculated to be $N_{I}^{0}=1.48 (\pm 0.43) \times 10^5$.

\tab The volume within which the ions are trapped, $V_{IT}$, is determined for the present trap parameters by a combination of trap loss and trap secular frequency measurements. A Monte-Carlo analysis of the time for the loss of ions from the ion trap~\cite{Rav12} gives the secular trap depth for trapped ions to be $\approx$ 0.8 eV. The secular frequencies for the ion trap are $\omega_x = \omega_y = 2 \pi \times 135 $ kHz, and $\omega_z = 2 \pi \times 27 $ kHz, which allow us to calculate the trap extent in each direction from the relation $ m_I \omega_r^2 r^2/2 = 0.8$ eV, where $r\in{x_0,y_0,z_0}$ is the extremal displacement in each direction and $m_I$ is the mass of the ion. The trapping volume defined by these dimensions for a single ion then is $V_{IT} = 8.3 (\pm 0.83) \times 10^{-8}$ m$^3$. 

\tab The collision rate is determined by substituting the mean values of the quantities above in eqn.~(\ref{eq:ion_atom_int_param}), to be
%------------------------------------------
\begin{equation}
\label{eq:collrate}
k_{ia} =  (1.23 \pm {0.42}) \times 10^{-13}~m^3/s.
\end{equation}
%-----------------------------------------
The standard deviation error is $\sigma_{k_{ia}} \equiv \sqrt{\sum_{i} \sigma_i^2} = 0.42 \times 10^{-13}$ m$^3$/s, where $\sigma_i$'s are the individual errors for each contributing parameter, shown in Table.~\ref{table:error}. This demonstrates the technique's ability to arrive at the rate coefficients for ion atom processes, even when the ions are not directly detected. The above rate coefficient is determined when the fraction of atoms in the excited state for $f_e \approx 0.28$, from eqn.~\ref{eq:exc_population}. The measured rate coefficient incorporates the elastic and resonant charge exchange collision from the ground and excited state of the atoms.

\subsection{Theoretical estimate of the collision rate coefficient}
\tab To estimate the collision rate coefficient in eqn.~(\ref{eq:ratecoeff}) we adopt the analytical form of total cross section as a function of collision velocity from ~\cite{Cot00, Mai07}, using the ground state $C_4$ value for Rb atoms, 5.26$\times 10^{-39}$Cm$^2$/V ~\cite{Ste12}. The speed distribution of trapped ions is determined using Monte-Carlo analysis and molecular dynamics simulations~\cite{Rav12}. Molecular dynamics simulation shows that the ion trap can stably trap ions with Maxwell Boltzman(MB) distribution of temperature $\approx$ 1000K. In Monte Carlo analysis, a non interacting distribution of ions are evolved in the absence of cold atoms in the ion trap potential to estimate the maximum trappable secular energy, which is found to be $\approx$ 0.8 eV, corresponding to speed of ~1360m/s. This value matches the tail of the MB distribution of ions in the trap. Since $V_{IT} \gg V_{MOT}$, and the two traps are well centered, in the overlap region, the micro motion velocity is small. Therefore, to estimate the speed distribution of the ions, only considering the secular velocity is a good approximation. By performing the integration in eqn.~(\ref{eq:ratecoeff}) with these quantities, we calculate the rate coefficient as 9.4$\times 10^{-14}$ m$^3$/s, when all the Rb atoms are in the ground state. Because atoms in the MOT are constantly pumped to the excited state by the cooling laser, ions invariably collide with a fraction of the MOT atoms, which are in the excited state atoms. In the experiment, we have 28$\%$ of atom population in the excited state ($f_e = 0.28$). The scalar polarizability for a Rb atom in the excited 5$p_{3/2}$ state is 14.15$\times 10^{-39}$Cm$^2$/V~\cite{Ste12}, which is larger than that in the ground state. Incorporating this fraction of atoms in the excited state in the calculation, we compute the rate coefficient as $k_{ia} = 1.12 \times 10^{-13}$ m$^3$/s, which agrees well with the experimental value.

\section{Discussion}
\tab The above technique allows the detection of collisional processes between trapped ions and atoms. In the present case, this is successful even for optically dark ions. At its core it relies on the measured changes in the fluorescence of the MOT atoms in the presence of the ions. The present work relies on two important and reasonable premises, which need to be emphasized. 

\tab The first is that ion-atom collisions are two body processes. This is reasonable because of several conditions that exist in the experiment. The atoms in the MOT are non-interacting to a good approximation. Because the ions are hot, they have large velocities and in this regime the binary ion-atom cross section is small. This is enough to make the presence of another ion in the vicinity of the collision, such that it affects the details of the ion-atom collision, highly improbable. It should be noted though that the experiment and the model are in good agreement, and since the model only incorporates two particle ion-atom collision, it is fair to argue that any more complex process is absorbed within the errors quoted in the present result.

\tab The second premise is that ion-atom collisions knock out a MOT atom. In the experiment, the only cooling mechanism for the ions is by collision with the cold atoms~\cite{Rav12}. The ions are also subject to significant continuous RF heating as their numbers are large and $V_{IT}\gg V_{MOT}$, while the cooling is most effective only at the center of the MOT. Thus the ions are quite hot and therefore possess large velocities. Under such circumstances, even a glancing ion-atom collision, whether elastic or resonant charge exchange, will transfer sufficient energy to the laser cooled atom such that it exceeds the capture velocity of the MOT and gets ejected from the MOT. This then forms the basis for equating the MOT loss rate with the ion-atom collision rate as described above.    

\tab Since the majority of ion-atom combination experiments use visible, laser cooled ions, we provide a brief discussion of some major differences. In such cases, since ion temperatures are easily in the mK levels, the ion-atom cross sections are much higher. The ions then would be crystallized and have a velocity distribution that would be well characterized. How atoms interact with such crystallized ions and what part the long range order of the ions plays in the measurement of the binary ion-atom interaction needs to be carefully understood. A possible problem is that an ion-atom collision at these energies may not result in the ejection of the colliding atom from the MOT and would therefore require eq.~(\ref{eq:collrateIA}) to be written with a proportionality constant. However if such an equation can be written, the cross-section could be directly determined because of the well characterized velocity and density distributions of the ions. Obviously, the rate equations constructed above would need to be modified according to the specifics of such a system.

\section{Conclusion}
\tab In a trapped ion and atom mixture, where the atoms are laser cooled and contained in a MOT and the ions are trapped within a Paul trap, we have developed a technique which measures the rate coefficient of ion atom collisions, even when the ions are optically dark. The rate equation formalism has been systematically developed for this experimental system. The experimental results are then fit to the rate equation model, which is seen to provide a consistent and adequate description for the measurements. Analysis of the experimentally measured quantities with the model allows the determination of the trapped Rb$^+$ ions and Rb atom collision rate coefficient. The value of the rate coefficient compares well with the theoretical estimation and validates the technique. 

\tab
\section{Acknowledgment}
\tab The authors gratefully acknowledge comments and suggestions made by Prof. G. Werth.
%%%%%%%%%%%%%%%%%%%%%%%%%%%%%%%%%%%%%%%%%%%%%%%%%%%%%%%%%%%%%%%%%%%%%%%%%%%%%%%%%%%%%%%%%%%%%%%%%%


\begin{thebibliography}{99}

\bibitem{Zip10}
C.\ Zipkes, S.\ Palzer, C.\ Sias, M.\ K\"{o}hl, Nature {\bf 464}, 388 (2010) 

\bibitem{Sch10}
S.\ Schmid, A.\ H\"{a}rter and J.\ H.\ Denschlag, Phys. Rev. Lett. {\bf 105}, 133202 (2010)

\bibitem{Gri09}
A.\ T.\ Grier, M.\ Cetina, F.\ Oru\v{c}evi\'{c}, and V.\ Vuleti\'{c}, Phys. Rev. Lett. {\bf 102}, 223201 (2009)

\bibitem{Fel11}
F.\ H.\ J.\ Hall, M.\ Aymar, N.\ Bouloufa-Maafa, O.\ Dulieu, and S.\ Willitsch,  Phys. Rev. Lett. {\bf 107}, 243202 (2011)

\bibitem{Wad11}
W.\ G.\ Rellergert, S.\ T.\ Sullivan, S.\ Kotochigova, A.\ Petrov, K.\ Chen, S.\ J.\ Schowalter, and E.\ R.\ Hudson,  Phys. Rev. Lett. {\bf 107} 243201 (2011)

\bibitem{Rav12}
K.\ Ravi, S.\ Lee, A.\ Sharma, G.\ Werth and S.\ A.\ Rangwala, Nat. Commun. 3:1126 doi: 10.1038/ncomms2131  (2012)

\bibitem{Sco11}
S.\ T.\ Sullivan, W.\ G.\ Rellergert, S.\ Kotochigova, K.\ Chen, S.\ J.\ Schowaltera and E.\ R.\ Hudson, Phys. Chem. Chem. Phys., {\bf 13} 18859 (2011)

\bibitem{Arn12}
A.\ H\"{a}rter, A.\ Kr\"{u}kow, A.\ Brunner, W.\ Schnitzler, S.\ Schmid, and J.\ H.\ Denschlag, Phys. Rev. Lett. {\bf 109}, 123201 (2012)

\bibitem{Fel12}
F.\ H.\ J.\ Hall and S.\ Willitsch, Phys. Rev. Lett. {\bf 109}, 233202 (2012)

\bibitem{Chu93}
D.\ A.\ Church, Physics Reports {\bf 228}, Nos. 5 and 6,  253-358 (1993)

\bibitem{Dra05}
G.\ W.\ F.\ Drake, Handbook of Atomic, Molecular and Optical Physics, Springer (2005)

\bibitem{Rav11}
K.\ Ravi, S.\ Lee, A.\ Sharma, G.\ Werth, and S.\ A.\ Rangwala, Appl. Phys. B {\bf 107}, 971 (2012)
doi 10.1007/s00340-011-4726-6 (2011)

\bibitem{Ste12}
D.\ A.\ Steck, Rubidium 85 D Line Data, Revision (2.1.5) (2012)

\bibitem{Maj05}
F.\ G.\ Major, V.\ N.\ Gheorghe, G.\ Werth, Charged Particle Traps, Springer (2005)

\bibitem{Tow95} 
C.\ G.\ Townsend, N.\ H.\ Edwards, C.\ J.\ Cooper, K.\ P.\ Zetie, C.\ J.\ Foot, A.\ M.\ Steane, P.\ Szriftgiser, H.\ Perrin, and J.\ Dalibard, Phys. Rev. A {\bf 52}, 1423 (1995)

\bibitem{San05}
J.\ E.\ Sansonetti, W.\ C.\ Martin, Handbook of Basic Atomic Spectroscopic data, American Institute of Physics (2005)

\bibitem{Gab97}
C.\ Gabbanini, S.\ Gozzini, and A.\ Luchesini, Op. Comm. {\bf 141}, 25-28 (1997)

\bibitem{Fus00}
F.\ Fuso, D.\ Ciampini, E.\ Arimondo, C.\ Gabbanini, Op. Comm.  {\bf 173}, 223-232 (2000) 

\bibitem{Cot00} 
R.\ C\^{o}t\'{e}, A.\ Dalgarno, Phys. Rev. A {\bf 62}, 012709 (2000)

\bibitem{Mai07}
S.\ A.\ Maiorov, O.\ F.\ Petrov, and V.\ E.\ Fortov, Calculation of resonant charge exchange cross sections of ions of
rubidium, cesium, mercury, and noble gases, 34th EPS Conf. on Plasma Phys., Warsaw, July 2007, ECA Vol. 31F, P-2.115 (2007).

\end{thebibliography}
\end{document}